\documentclass[%
 reprint,
superscriptaddress,
 amsmath,amssymb,
 aps,
]{revtex4-2}

\usepackage{graphicx}
\usepackage{dcolumn}
\usepackage{lipsum}
\usepackage{bm}
\usepackage{tikz}
\usepackage{siunitx}
\usepackage{svg}
\usepackage{appendix}
\usepackage{flushend} 
\usepackage{url}
\usetikzlibrary{quantikz2} 

\newcommand{\ketbra}[2]{|#1\rangle\langle#2|}

\newcommand{\kett}[1]{$\left|#1\right\rangle$}
\renewcommand{\bibsection}{%
    
    \onecolumngrid
  \vspace{2em}
  \noindent\hbox to \textwidth{\hfil\vrule width 0.5\textwidth height 0.4pt\hfil}
  \vskip 1em
  \twocolumngrid
}

\begin{document}
\makeatletter

\preprint{}

\title{Quantum Fanout Gates in Constant Depth via Resonance Engineering}

\author{Johannes Alexander Jaeger}

\email{aljaeger@ethz.ch}
\affiliation{Institute for Quantum Electronics, ETH Z\"urich, 8093 Z\"urich, Switzerland
}%
\author{Elias Zapusek}%
\affiliation{Institute for Quantum Electronics, ETH Z\"urich, 8093 Z\"urich, Switzerland
}%
\author{Florentin Reiter}%

\affiliation{Institute for Quantum Electronics, ETH Z\"urich, 8093 Z\"urich, Switzerland
}
\affiliation{Fraunhofer  Institute for Applied Solid State Physics IAF, Tullastr. 72, 
79108 Freiburg, Germany}%


\date{\today}

\begin{abstract}
We present a novel implementation of an $n$-qubit fanout gate using resonance engineering. Our proposed mechanism uses Jaynes-Cummings interactions between multiple qubits and a common harmonic oscillator to realize a fanout gate at the system-level. Our theoretical analysis establishes upper bounds on the gate error, demonstrating linear infidelity scaling in constant time -- a favorable trade-off compared to a conventional CNOT decomposition. To validate the performance of our scheme at large system sizes, we exploit permutation symmetry to reduce the simulation complexity from exponential to polynomial in the number of qubits, enabling simulation up to 100 qubits. The results of this numerical analysis are consistent with our theoretical findings and allow us to characterize the performance well. Our gate will enable faster stabilizer readouts and could provide polynomial speedups in many quantum algorithms.

\end{abstract}

\maketitle


\section{Introduction}
Scaling quantum computation to useful problem sizes requires not only the control of a large number of qubits, but also the implementation of efficient many-qubit operations \cite{preskillQuantumComputingNISQ2018}. Although universal quantum computation can be realized with sequences of two-qubit gates \cite{barencoElementaryGatesQuantum1995,nielsenQuantumComputationQuantum2012}, decomposing complex operations into elementary gates incurs significant circuit depth and error accumulation. The implementation of system-level many-qubit gates can offer practical speedups for quantum computation, motivating ongoing research into novel gate constructions that can improve both fidelity and scalability \cite{mooreQuantumCircuitsFanout1999,hoyerQuantumCircuitsUnbounded2005}.

The fanout gate, also known as the multi-target controlled-NOT gate, has been shown to provide significant speedups in many crucial quantum algorithms. Several arithmetic operations -- including modulo, parity and the approximate quantum Fourier transform \cite{coppersmithApproximateFourierTransform2002} -- can be implemented in constant depth with a fanout gate. This drastically reduces the circuit depth in many quantum subroutines, including syndrome measurements for quantum error correction \cite{shorSchemeReducingDecoherence1995, steaneErrorCorrectingCodes1996,fowlerSurfaceCodesPractical2012,baconOperatorQuantumErrorcorrecting2006,calderbankGoodQuantumErrorcorrecting1996}, the Grover search algorithm \cite{groverQuantumMechanicsHelps1997,hoyerQuantumCircuitsUnbounded2005} and quantum circuit sampling \cite{foxmanRandomUnitariesConstant2025}. The gate can be implemented transversally for many error correcting codes; if the controlled-NOT gate can be implemented transversally with system-level controlled-NOT gates, the fanout gate can be implemented transversally with system-level fanout gates. This is the case for many quantum error correcting codes, such as CSS codes \cite{gottesmanIntroductionQuantumError2009}.

\begin{figure}
    \centering
    \includegraphics[width=\linewidth]{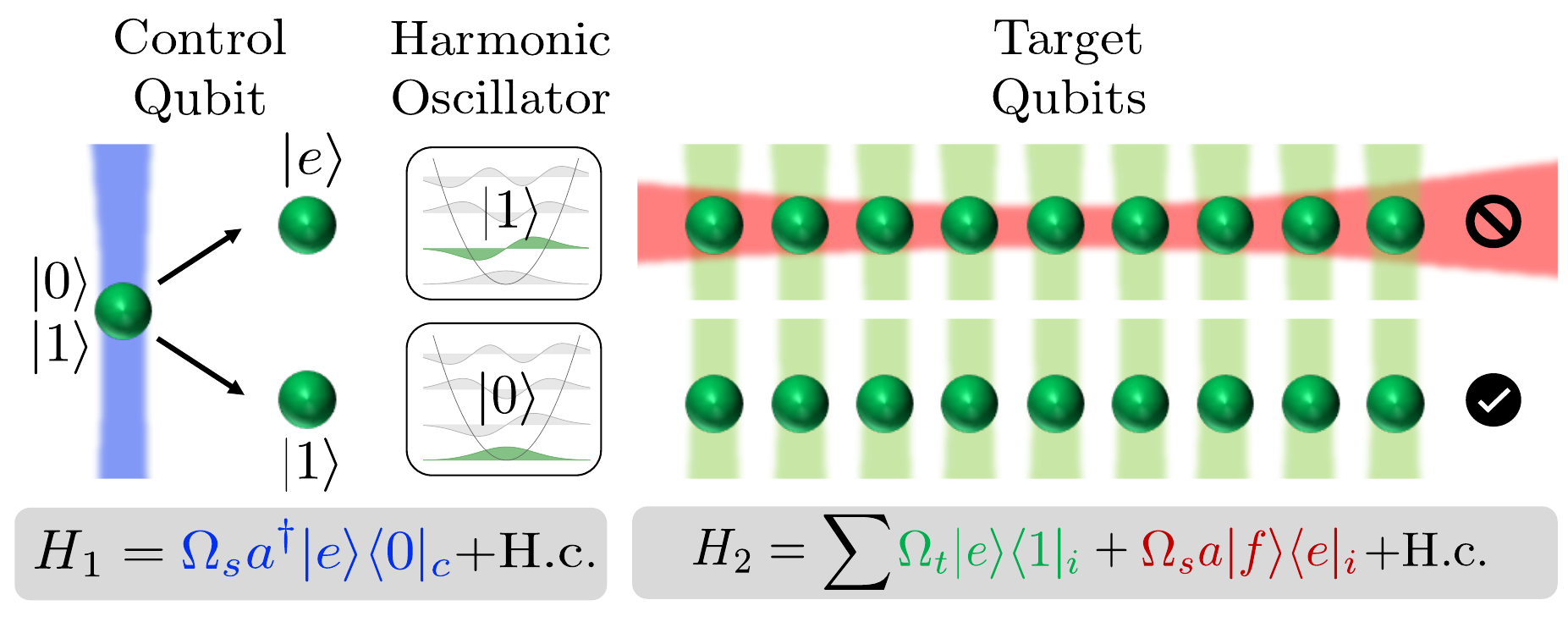}
    \caption{Schematic representation of the fanout gate scheme. The qubit state is transferred to the harmonic oscillator using a Jaynes-Cummings drive (blue). This allows for the selective blocking of the Rabi drives (green) on the target qubits using a strong Jaynes-Cummings drive (red).}
    \label{fig:fanout_schematic_beams}
\end{figure}

Most existing proposals for fanout rely on pairwise interactions between qubits, either through engineered direct couplings or effective power-law interactions on a lattice \cite{fennerImplementingQuantumFanout2023,fennerImplementingFanoutParity2004,guoImplementingFastUnbounded2022}. Other approaches have implemented fanout using measurements  \cite{songConstantdepthFanoutRealtime2025,baumerMeasurementbasedLongrangeEntangling2025}, which avoid deep unitary decompositions but are constrained by measurement fidelity and latency. By contrast, our method exploits the simultaneous Jaynes–Cummings coupling \cite{jaynesComparisonQuantumSemiclassical1963} of all qubits to a common harmonic oscillator mode to implement the fanout gate directly at the system level, providing a direct realization of a global multi-qubit operation in constant depth. This advantage carries over to quantum algorithms, allowing for a circuit depth reduction in implementations of many quantum algorithms which use the fanout gate as a subroutine. 

We will consider the implementation of a phase-flip fanout, which conditionally applies the phase gate $Z$ to multiple target qubits conditioned on one control qubit. This gate requires less drives to be implemented, and is fully equivalent to a bit-flip fanout up to local operations. We assume a system consisting of $4$-level systems, coupled to a common harmonic oscillator via (anti-) Jaynes-Cummings interactions 
\begin{equation}
    H_{JC} = g\left(a\sigma_+ + a^\dag \sigma_-\right )\text{,}
\end{equation}
where $a$ is the annihilation operator of the harmonic oscillator and $\sigma_\pm$ are the raising and lowering operator of the qubit. Such operations are common on many quantum platforms, such as trapped ions coupling to a motional mode, and superconducting circuits coupling to an optical resonator \cite{ciracQuantumComputationsCold1995,blaisCavityQuantumElectrodynamics2004,blaisCircuitQuantumElectrodynamics2021}.

We present a schematic representation of gate mechanism in Fig. \ref{fig:fanout_schematic_beams}. We first transfer the control qubit state to the harmonic oscillator. In a subsequent step, we apply a strong Jaynes-Cummings interaction on the target qubits which annihilates an excitation on the harmonic oscillator. A second (weaker) drive will be blocked by this strong interaction in the same sense as electromagnetically induced transparency (EIT) \cite{marangosElectromagneticallyInducedTransparency1998,fleischhauerElectromagneticallyInducedTransparency2005}. Since the Jaynes-Cummings drive requires an excitation on the harmonic oscillator to couple, this allows us to implement conditional operations. 

We discuss our scheme in depth, both analytically and numerically. We begin by presenting an overview of the technique of resonance engineering, which we use to engineer the fanout gate, in Sec. \ref{sec:resonanceengineering}. After outlining the requirements of our system in Sec. \ref{sec:system}, we introduce the gate mechanism in Sec. \ref{sec:gatemechanism}. We analyze the scheme in Sec. \ref{sec:theoreticalanalysis}, deriving analytic fidelity bounds of our method which scale linearly with the number of qubits. In the subsequent numerical analysis of Sec. \ref{sec:numericalanalysis}, we verify the validity of the theory with simulations up to 100 qubits and analyze the effect of heating on the gate fidelity.

\section{Resonance Engineering}\label{sec:resonanceengineering}
Resonance is a fundamental concept in physics, appearing in many different contexts ranging from acoustics to particle physics. The canonical example of resonance is the classical harmonic oscillator; If a harmonic oscillator is subjected to a periodic external driving force, the harmonic oscillator can be driven to higher amplitudes. One finds that if one drives at the right frequency, corresponding to the eigenfrequency of this oscillator, the excitation amplitude is maximal. This is known as resonance. If we introduce a coupling in a harmonic oscillator, the system forms new normal modes, which generally have a shifted resonance frequency. 

This concept transfers analogously to quantum systems. If we have a quantum system and introduce an interaction, we can consider the eigenstates of the system including the interaction Hamiltonian. These states are referred to as \textit{dressed states}, which generally have a shifted energy. This energy shift can be engineered to change the resonance properties of a system, which we leverage as a tool to implement entangling operations. This method is referred to as \textit{resonance engineering} (also Hilbert space engineering), and has been used successfully for entangled state preparation \cite{linPreparationEntangledStates2016} and the implementation of non-unitary gates \cite{vanmourikExperimentalRealizationNonunitary2024}. 

Suppose we describe a system by a time independent Hamiltonian $H$. We can divide the system as
\begin{equation}
    H = H_c + H_p\text{,}
\end{equation}
where $H_c$ is a drive which we use to change the eigenstates of the system, and $H_p$ is used to drive transitions between the new eigenstates.

The \textit{dressed state basis} for the drive $H_c$ is an eigenbasis of $H_c$. In the dressed state basis, the effect of $H_c$ is simply an energy shift of the dressed states. If we express $H_p$ in this basis, we find that a drive that was previously resonant may no longer be resonant in the dressed state picture, due to the energy shift of the dressed states. Similarly, one can choose an off-resonant drive which becomes resonant with one of the dressed states after the interaction $H_c$ is introduced.

\begin{figure}[h]
    \centering
    
    \includegraphics[width=0.6\linewidth]{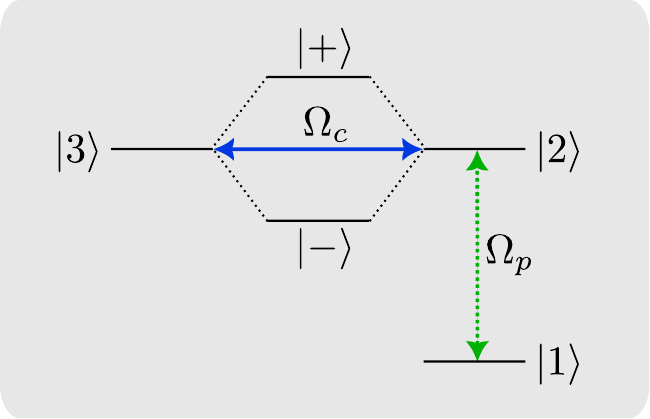}
    \caption{An energy level diagram for the EIT system introduced in Eq. \eqref{eq:EIT}. The blue drive couples the states $\ket 2$ and $\ket 3$ and is assumed to be far stronger than the green drive coupling the states $\ket 1$ and $\ket 2$. We consider the system in the dressed state picture of the coupling drive $\Omega_c$, under which the states $\ket 2$ and $\ket 3$ hybridize to form dressed states $\ket \pm = \frac{\ket{2} \pm \ket 3}{\sqrt 2}$ with energy shift $E_\pm = \pm \Omega_c$, as given by the Hamiltonian in Eq. \eqref{eq:EIT_dressed}. In this picture, the excited states have a shifted energy, making the probe drive off-resonant and preventing absorption.}
    \label{fig:resonance_engineering}
\end{figure}

\medskip
As an example, one can view electromagnetically induced transparency (EIT)  through the framework of resonance engineering. EIT is the effect that one laser drive can be used to prevent the absorption of another, making the material ``transparent'' at the corresponding frequency \cite{marangosElectromagneticallyInducedTransparency1998}. The typical EIT system is shown in Fig. \ref{fig:resonance_engineering}. It is presented here because our proposed gate implementation follows a similar Hamiltonian.

Consider three states $\{\ket 1, \ket 2, \ket 3\}$, coupled by two drives which we refer to as the coupling drive $\Omega_c$ and the probe drive $\Omega_p$, where we assume $|\Omega_c|^2 \gg |\Omega_p|^2$. We define the Hamiltonian of the system as
\begin{equation}\label{eq:EIT}
    H = \Omega_c\ketbra{2}{3} + \Omega_p \ketbra{1}{2} + \text{H.c.} \text{,}
\end{equation}
where we identify
\begin{equation}
    H_c = \Omega_c\ketbra{2}{3} + \text{H.c.} \quad \text{and} \quad H_p = \Omega_p \ketbra{1}{2} + \text{H.c.}
\end{equation}
The drive $H_c$ has dressed states given by
\begin{align}
    \ket{\pm} = \frac{\ket{2} \pm \ket{3}}{\sqrt 2} \quad \text{with energy shift } E_\pm = \pm \Omega_s\text{.}
\end{align}
Expressing the Hamiltonian in the dressed state basis then results in
\begin{equation}\label{eq:EIT_dressed}
    H = \Omega_s \left(\ketbra{+}{+} - \ketbra{-}{-}\right) + \frac{\Omega_p}{\sqrt 2} \left(\ketbra{1}{+} + \ketbra{1}{-} + \text{H.c.}\right)
\end{equation}

In this basis, we have two strongly off-resonant drives on the state $\ket 1$, and hence we find that absorption due to the probe drive is suppressed. This results in electromagnetically induced transparency, which we derived here using the resonance engineering picture. In the dressed state picture, the probe drive splits into two off-resonant drives via two auxiliary levels, with opposite detuning. It turns out that the AC stark shifts of these two drives cancel, as is demonstrated in App. \ref{sec:cancelling_raman}.

In Sec. \ref{sec:gatemechanism}, we introduce our gate mechanism for the fanout gate, which leverages resonance engineering to conditionally block a transition based on the oscillator state. We find that on the subspaces where the transition is blocked, the Hamiltonian has the exact same form as the above EIT system.

\section{System}
\label{sec:system}

Our implementation requires qubits with two auxiliary levels. We will denote these levels with the basis states
\begin{equation}
    \ket 0, \ket 1, \ket e \text{ and} \ket f\text{,}
\end{equation}
where $\ket 0$ and $\ket 1$ are referred to as computational basis states and $\ket e$ and $\ket f$ excited states.

We will further assume that each qubit can be coupled to a cold common harmonic oscillator, which we denote by the annihilation operator
\begin{equation}
    a = \sum_{n=0}^\infty \sqrt{n+1} \ketbra{n}{n+1}\text{,}
\end{equation}
where $\ket n$ are the different energy eigenstates of the harmonic oscillator.

We will assume that we can implement local operations on the qubits, given by
\begin{equation}
    H = \Omega (\ketbra{j}{k}_i + \ketbra{k}{j}_i)\text{,}
\end{equation}
as well as Jaynes-Cummings type interactions with the common harmonic oscillator given by
\begin{equation}
    H = \Omega_s (a\ketbra{j}{k}_i + a^\dag\ketbra{k}{j}_i)\text{,}
\end{equation}
where $k,j \in \{0,1,e,f\}$ and $i$ is the index corresponding to the qubit. Note that in order to implement our gate scheme, the control and target qubits must be addressed separately.

The gate mechanism proposed in this article is platform agnostic; it can be used on any quantum computation platform provided the above interactions can be implemented.

These interactions have been demonstrated on various quantum platforms. In trapped ions, the Jaynes-Cummings interaction above could be implemented as a sideband transition between the ions internal degrees of freedom and a collective motional mode in the Lamb-Dicke limit \cite{leibfriedQuantumDynamicsSingle2003}. In superconducting circuits, Jaynes-Cummings interactions have been demonstrated between transmon qubits and a micromachined cavity resonator \cite{brechtMicromachinedIntegratedQuantum2017}.

\section{Gate Mechanism}\label{sec:gatemechanism}
We focus our attention to the implementation of the phase-flip fanout gate, which applies conditional $Z$ gates instead of $X$ gates. This is typically easier to implement as it only requires a conditional phase shift on computational basis states, rather than population transfer. These gates are fully equivalent up to local transformations and are thus equally powerful. We generate this phase shift by applying a $2\pi$ pulse to the $\ket 1$ state through an auxiliary state. We will then use resonance engineering to conditionally block this state transfer. To make this operation conditional, we will first excite the harmonic oscillator conditioned on the control qubit using a Jaynes-Cummings drive, and then apply resonance engineering to block a transition on the one-phonon manifold. The gate sequence is shown below.

\begin{figure}
    \centering
    \includegraphics[width=\linewidth]{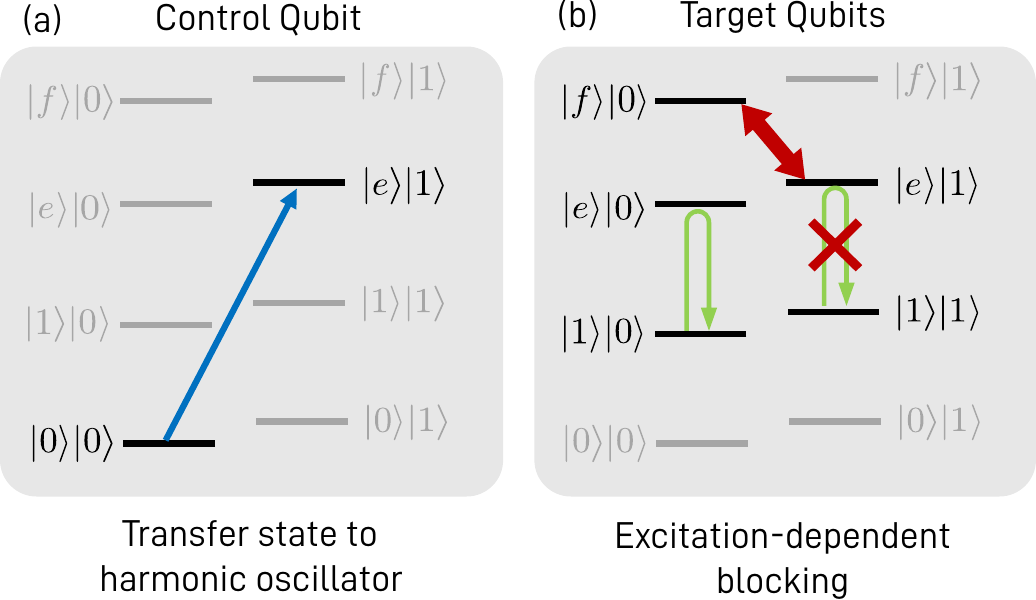} 
    \caption{An energy level depiction of the steps of the gate mechanism. The states $\ket{a}\ket{b}$ represent the tensor product of a qubit state $\ket{a}$ and HO state $\ket{b}$. Panel (a) depicts the blue sideband $\pi$-pulse on the control qubit used to generate or remove an excitation. Panel (b) shows how resonance engineering is used to apply a phase to the $\ket 1$  state on the target qubits conditioned on the number of excitations. A strong red sideband is applied to the target qubit, from an auxiliary state $\ket e$ to a second auxiliary state $\ket f$. As this is a red sideband, it annihilates a phonon on the $\ket e$ state and therefore only couples to the 1-phonon state, while the 0-phonon state is dark. In the dressed state picture, this causes a phonon-dependent shift on the state $\ket e$. We then add a $2\pi$ pulse carrier drive (green) from the $\ket 1$ state to the $\ket e$ state. On the 0-phonon manifold, the drive is resonant and implements a $Z$ gate. On the 1-phonon manifold, this forms a canceling off-resonant drive and is blocked (see App. \ref{sec:cancelling_raman}).}
    \label{fig:4-level-pulsed}
\end{figure}
\medskip
\textbf{Control Sequence:}
\begin{enumerate}
    \item Excite the harmonic oscillator conditioned on the control qubit using a Jaynes-Cummings $\pi$-pulse:
    \begin{equation}
        H_1 = \Omega_c a^\dag\ketbra{e}{0}_1 + \text{H.c.}
    \end{equation}
    \item Apply conditional phase to target qubits using a $2\pi$ pulse carrier drive $\Omega_t$, while applying a strong anti-Jaynes-Cummings drive $\Omega_c$ to block the transition on the one phonon manifold:
    \begin{equation}
        H_2 = \sum_{i=2}^{n} \Omega_c a\ketbra{f}{e}_i + \Omega_t \ketbra{e}{1}+\text{H.c.} \label{eq:H_2}
    \end{equation}
    \item Restore the control qubit state and de-excite the harmonic oscillator using a Jaynes-Cummings $\pi$-pulse:
    \begin{equation}
        H_3 = \Omega_c a^\dag\ketbra{e}{0}_1 + \text{H.c.}
    \end{equation}
\end{enumerate}

The steps of the control sequence are shown on an energy level diagram in Fig. \ref{fig:4-level-pulsed}. Here it is easier to see how the drives operate on the different subspaces. In the first step (depicted in panel  (a)), the harmonic oscillator is excited if the control qubit is in the state $\ket{0}$. In the second step, the presence (or absence) of this excitation will change how the Hamiltonian operates (see Fig. \ref{fig:4-level-pulsed}b).

Without the excitation, the drive $\Omega_c$ of $H_2$ is dark on all states except the \kett{f} state. Since the probe drive acts between $\ket{1}$ and $\ket{e}$, $H_2$ simply applies a $2\pi$ pulse through an excited state on all target qubits, and implements the phase $Z^{\otimes n-1}$ exactly. 

In the presence of an excitation on the harmonic oscillator, the anti-Jaynes-Cummings drive between $\ket{e}$ and $\ket{f}$ couples to the state \kett{e}, blocking the transition driven by the probe drive (depicted in green in Fig. \ref{fig:4-level-pulsed}b). Note that this system is described by a Hamiltonian of the same form as the EIT example discussed in the introductory sections (see Fig. \ref{fig:resonance_engineering}). 

The astute reader may wonder how this blocking of the probe drive generalizes to many target qubits. We find that the dynamics are analogous to the case of one target qubit, by performing a change of basis on the excited states. We will demonstrate this in the theoretical analysis below.

\section{Theoretical Analysis}\label{sec:theoreticalanalysis}
In this section we will derive an analytic lower bound for the fidelity of this fan-out gate implementation. 

The average gate fidelity is given by
\begin{equation}
    F = \int  d\psi |\bra{\psi} U^\dag e^{-i H t} \ket{\psi}|^2 \text{,}
\end{equation}
where $U$ is the ideal gate operation and $H$ is the generating Hamiltonian of our system dynamics.

We approximate this as
\begin{equation}\label{eq:approx_fidelity}
    F = \frac{1}{|B|}\sum_{b\in B}  \left|\bra{b} U^\dag e^{-i H t} \ket{b}\right|^2\text{.}
\end{equation}
This approximation is only valid under the assumption that the Hamiltonian does not mix basis vectors. We therefore search to find a basis for which $B$ for which $U^\dag e^{-i H t}$ is diagonal.

To evaluate the fidelity, we consider two subspaces: The idle subspace (control qubit $\ket 0$) and the transition subspace (control qubit $\ket 1$).

\subsubsection{Transition Subspace $\ket 1_c$}
The transition subspace is not excited by the $\pi$ pulse in the first step of Fig. \ref{fig:4-level-pulsed}, and therefore the harmonic oscillator remains in the ground state. In the second step, we apply a carrier drive $\ket{1}$ to $\ket{e}$ and an anti-Jaynes-Cummings drive $\ket{e}$ to $\ket{f}$ to the target qubits. Since the anti-Jaynes-Cummings drive needs to annihilate an excitation, it does not operate on the state $\ket{1}$ if we are in the ground state. This leaves only the $2\pi$ pulse implemented by the carrier drive, therefore implementing an ideal $Z$ gate on the target qubits. In the absence of imperfections, this implements the operation 
$$\ketbra{1}{1}\otimes Z^{n-1}$$
ideally, and therefore results in fidelity $1$.

\subsubsection{Idle Subspace $\ket 0_c$}
\begin{figure}[h]
    \centering
    \includegraphics[width=\linewidth]{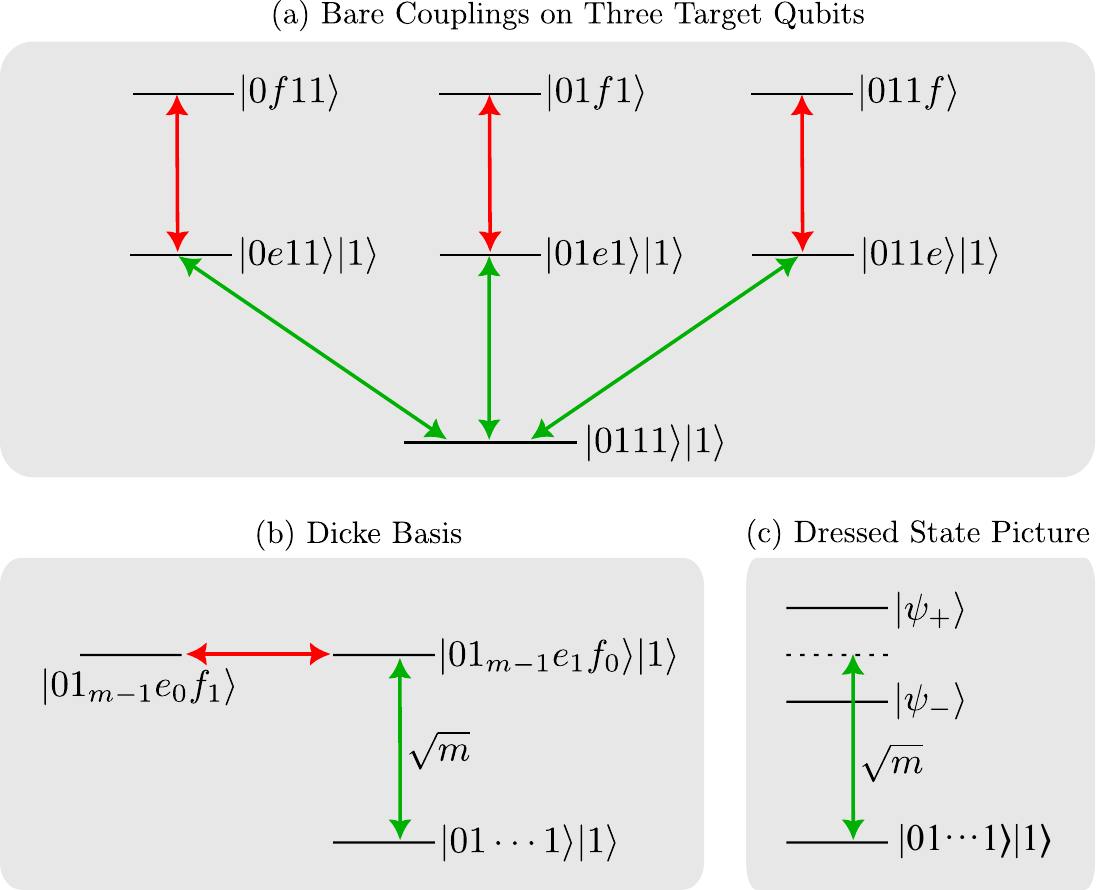}
    \caption{Gate dynamics for excitation-dependent blocking of the probe drive for the case of multiple target qubits. (a) Couplings for three target qubits in the tensor-product basis, with higher-order excitations induced by the probe drive (green) neglected, as resonance engineering suppresses population transfer out of the ground manifold.
    (b) The couplings expressed in terms of Dicke states, generalized for a state with $m$ target qubits in the $\ket 1$ state. The notation $\ket{01_{n_1}e_{n_e}f_{n_f}}$ is an equal superposition of states with control qubit $0$, $n_e$ target qubits in the $\ket e$ state, and so on. 
    (c) The same interaction as panel (b) in the dressed state picture of the red drive.  This forms canceling off-resonant drives, as discussed in Sec. \ref{sec:cancelling_raman}, which have no effective AC-stark shift on the computational state, thereby introducing no phase on the target qubit state. This system therefore approximately implements the identity on the computational subspace, aside from off-resonant excitations into the excited state manifold.}
    \label{fig:4-level-phonon_combined}
\end{figure}

If the control qubit is in the $\ket 0_c$ state, the first step will excite the harmonic oscillator to the first fock state. On the excited manifold, a state with $m$ target qubits in the $\ket 1$ state couples to $m$ excited states, the states with one target qubit in the $\ket e$ state. Each of these states is coupled further via the red sideband to a state with one qubit in the $\ket f$ state while removing an excitation from the harmonic oscillator. This is shown for the special case of three target qubits in Fig. \ref{fig:4-level-phonon_combined} (a). 

The red sideband will create dressed states that will shift the carrier probe drive off-resonant, as we will demonstrate below. 

We can express this interaction more simply in terms of Dicke states. We introduce the notation $\ket{c1_{n_1}e_{n_e}f_{n_f}}$ to denote an equal superposition of states with control qubit state $\ket c$, $n_1$ target qubits in the $\ket 1$ state, $n_{e}$ target qubits in the $\ket e$ state, etc. In this basis, the couplings in panel (a) of Fig. \ref{fig:4-level-phonon_combined} can be expressed in terms of collective couplings to the Dicke states, as shown in Fig. \ref{fig:4-level-phonon_combined} (b). This results in a collective enhancement of $\sqrt m $. Specifically, we notice that each computational basis state couples to different auxiliary states, and lies in different blocks of the Hamiltonian. We can therefore compute their evolutions separately, as we will do below. 

The Hamiltonian coupling to a state with $m$ qubits in the $\ket 1$ state is given by
\begin{align}
    H_m =& \Omega_c \ketbra{01_{m-1}e_1f_0, 1}{01_{m-1}e_0f_1} \nonumber \\
    &+ \Omega_t\sqrt m\ketbra{01_{m-1}e_1f_0, 1}{01\cdots 1} +\text{H.c.}
\end{align}

The dressed states of the drive $\Omega_c$ are given by
\begin{align}
    \ket{\psi_\pm} = \frac{\ket{01_{m-1}e_1f_0}\ket{1} \pm \ket{01_{m-1}e_0f_1}}{\sqrt 2}\text{,}
\end{align}
with energy shift $E_\pm = \pm\Omega_c$.

In this basis, the corresponding Hamiltonian is
\begin{align}
    H_m &= \Omega_c \Big(\ketbra{\psi_+}{\psi_+} - \ketbra{\psi_-}{\psi_-}\Big) \nonumber\\ 
    &+ \frac{\Omega_t}{\sqrt 2}\sqrt{m} \Big(\ketbra{\psi_+}{0b_1\cdots b_{n-1}} + \ketbra{\psi_-}{0b_1\cdots b_{n-1}} \nonumber\\ 
    &+ \text{H.c.}\Big)
\end{align}
The corresponding energy level diagram is shown in panel (c) of Fig. \ref{fig:4-level-phonon_combined}.

The off-resonant drives cause the state to remain idle, up to small off-resonant excitations out of the computational subspace. This can be seen from the idle probability computed in App. \ref{sec:cancelling_raman} given by 
\begin{align}
    &|\bra{0b_1\cdots b_{n-1}} \exp(-iH_m t)\ket{0b_1\cdots b_{n-1}}|^2 \nonumber\\
    &= \left(1-\frac{2m\Omega_t^2}{\Omega_c^2 + m\Omega_t^2}\sin^2\left(\frac{t}{2}\sqrt{\Omega_c^2 + m\Omega_t^2}\right)\right)^2 \nonumber\\
    &\approx 1-\frac{4m\Omega_t^2}{\Omega_c^2 + m\Omega_t^2}\sin^2\left(\frac{t}{2}\sqrt{\Omega_c^2 + m\Omega_t^2}\right) \text{,}
\end{align}
where we approximated using $ \Omega_t^2 \ll \Omega_c^2 $ in the last line. This idle probability encodes the probability to stay within the computational subspace, and can be averaged over all computational basis states to compute the average gate fidelity. We will do this in the following section.

\subsubsection{Fidelity}\label{sec:4-level-phonon-fidelity} 
Using the evolutions on the two subspaces, we can evaluate the fidelity using the approximate formula given in Eq. \eqref{eq:approx_fidelity}.

We then find a infidelity of 
\begin{align}
    1-F &= \frac{1}{2^n} \sum_b {1-|\bra{b} U^\dag e^{-i H t} \ket{b}|^2}\\
    & \leq\frac{1}{2^n}\sum_m \begin{pmatrix}
        n-1\\m
    \end{pmatrix} \frac{4m\Omega_t^2}{\Omega_c^2+m\Omega_t^2}\\
    &\leq \frac{1}{2^{n}} \frac{\Omega_t^2}{\Omega_c^2} \sum_m 4m\begin{pmatrix}
        n-1\\m
    \end{pmatrix} \text{.}
\end{align}

This sum can be evaluated to
\begin{equation}
     \frac{1}{2^n}\sum_{m=0}^n \begin{pmatrix}
        n\\m
    \end{pmatrix}m = \frac{n}{2} \text{,}
\end{equation}
for all $n\in \mathbb{N}$ using the Binomial theorem, as shown in App. \ref{sec:sum_calculation}. We therefore find the fidelity
\begin{equation}
    1-F\leq (n-1)\frac{\Omega_t^2}{\Omega_c^2}\text{.}
\end{equation}

Given that the gate duration is 
\begin{equation}
    t = \frac{\pi}{\Omega_t} \text{,}
\end{equation}
we can express the fidelity in terms of the gate duration as
\begin{equation}\label{eq:4-level-phonon-fidelitybound}
    1-F \leq (n-1) \frac{\pi^2}{\Omega_c^2 t^2}\text{.}
\end{equation}

If we wish to maintain a constant fidelity, we need a gate duration
\begin{equation}
    t = O\left(\sqrt n\right)\text{.}
\end{equation}

Note that a decomposition into $CZ$ gates requires a time $O(n)$ if they cannot be executed in parallel, and generally does not have a constant fidelity in the number of qubits. The different $CZ$ gates do not act on disjoint subspaces in the tensor product basis, and therefore adding more gates therefore results in higher infidelity. The exact fidelity scaling generally depends on the  tomography of the gate, but we expect it to scale approximately as
\begin{equation}
    1-F_n \approx (1-p)^n \approx 1- p n \text{,}
\end{equation}
where $p$ is the CZ gate infidelity, meaning that the gate error would also be $O(n)$ for a decomposition into CZ gates.

\subsubsection{Timing Condition}
\begin{figure}[h]
    \centering
    \includegraphics[width=\linewidth]{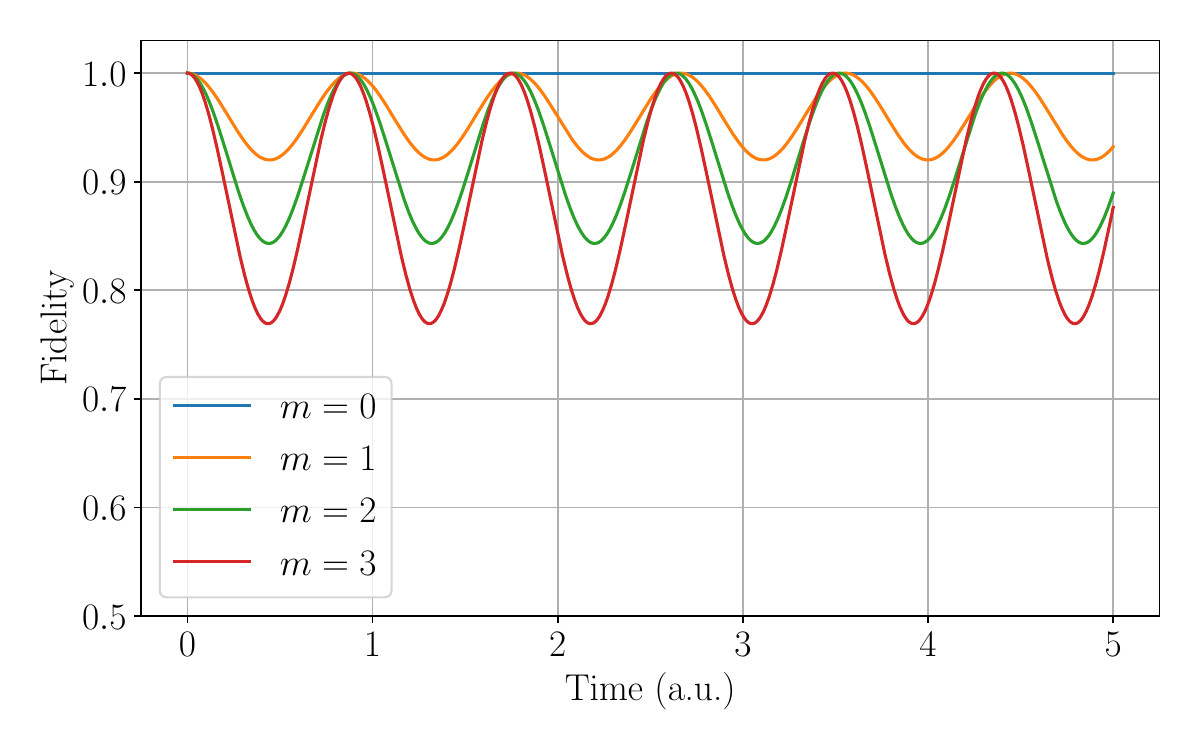}
    \caption{Fidelity of computational basis states with different number $m$ of target qubits in the state $\ket 1$ on the idle subspace (control qubit $\ket 0$), as a function of time. Notice that due to $\Omega_t^2 \ll \Omega_c^2$ (1 and 49 in the example), the off-resonant excitations occur at approximately the same frequency. }
    \label{fig:offres-excitations}
\end{figure}

The infidelity of the different computational basis states in the idle subspace is shown in Fig. \ref{fig:offres-excitations}. We notice that in the limit $\Omega_c^2  \gg m\Omega_t^2$, the off-resonant excitations all have approximately the same frequency. It is therefore possible to time the gate in such a way, that the off-resonant excitation disappears. 

 We will use this to derive an optimal timing condition which minimizes the off-resonant excitation. We established that the fidelity is given by
\begin{equation}\label{eq:fidelitytimingcondition1}
    1-F = \frac{1}{2^n}\sum_m \begin{pmatrix}
        n-1\\m
    \end{pmatrix} \frac{4m\Omega_t^2}{\Omega_c^2+m\Omega_t^2}\sin^2\left(\frac{t}{2}\sqrt{\Omega_c^2 + m\Omega_t^2}\right) \text{.}
\end{equation}

To cancel the off-resonant excitations, we want the value in the sine function to be approximately a multiple of $\pi$. To that end, we make the ansatz
\begin{equation}\label{eq:4lp_timing_condition}
    t = \frac{\pi}{\Omega_t} =2\pi k \frac{1}{\sqrt{\Omega_c^2 + m_\text{eff}\Omega_t^2}} \qquad \text{ where } k\in \mathbb{Z}\text{,}
\end{equation}
where $m_\text{eff}$ is a value to be determined which maximizes the fidelity. It can be thought of as an effective number of ones for the evolution, and we expect it to be somewhere around $\frac{n}{2}$.

The above condition can be rearranged to
\begin{equation}\label{eq:4lp-timing-condition}
    t = \frac{\pi}{\Omega_c} \sqrt{4k^2 - m_\text{eff}}\text{.}
\end{equation}

Substituting the timing condition into Eq. \eqref{eq:fidelitytimingcondition1}, we find
\begin{equation}
    1-F = \frac{1}{2^n}\sum_m \begin{pmatrix}
        n-1\\m
    \end{pmatrix} \frac{4m\Omega_t^2}{\Omega_c^2+m\Omega_t^2}\sin^2(\pi k\alpha) \text{,}
\end{equation}
where $\alpha = \sqrt{\left ({1 + m\frac{\Omega_t^2}{\Omega_c^2}}\right) \Big / \left( {1 + m_\text{eff}\frac{\Omega_t^2}{\Omega_c^2}}\right)}$.\\

Here it can be seen why we chose this ansatz: We find that in the limit $\Omega_t^2 / \Omega_c^2 \ll 1$, we find $\alpha \approx 1$. The value in the $\text{sin}$ function is therefore on average a multiple of $\pi$, resulting in an infidelity near zero. To quantify the remaining error, we perform a Taylor series in the variable $x=\frac{\Omega_t^2}{\Omega_c^2}$ to find
\begin{equation}
    1-F \approx \frac{1}{2^n}\sum_m \begin{pmatrix}
        n-1\\m
    \end{pmatrix} \pi^2 4 k^2 (m_\text{eff} - m)^2 m \frac{\Omega_t^6}{\Omega_c^6}\text{.}
\end{equation}
Evaluating the sum, we find
\begin{equation}
    1-F = \frac {\pi^2}{4} (n-1)\left(n-2 + (n-2m_\text{eff})^2\right) \frac{\Omega_t^6}{\Omega_c^6} k^2\text{.}
\end{equation}
We find terms quadratic in $n$, and cubic in $n$. However, by choosing $m_\text{eff} = \frac{n}{2}$, the cubic term cancels and we are left with
\begin{equation}
    1-F = \frac {\pi^2}{4} (n-1)(n-2) \frac{\Omega_t^6}{\Omega_c^6} k^2\text{.}
\end{equation}

It is important to note that $k$ depends on the drive strengths via the timing condition. If we approximate the timing condition (Eq. \eqref{eq:4lp_timing_condition}) to lowest order in $x = \frac{\Omega_t}{\Omega_c}$, it becomes
\begin{equation}
    t = \frac{\pi}{\Omega_t} = \frac{2\pi k}{\Omega_c} + O\left(\frac{\Omega_t}{\Omega_c}\right)
\end{equation}
which results in 
\begin{equation}
    k = \frac{\Omega_c}{2\Omega_t}\text{.}
\end{equation}

We therefore find 
\begin{equation}
    1-F = \frac {\pi^2}{16} (n-1)(n-2) \frac{\Omega_t^4}{\Omega_c^4}\text{.}
\end{equation}

Substituting in the timing condition $\Omega_t = \frac{\pi}{t}$ results in
\begin{equation}\label{eq:4-level-phonon-fidelitytimed}
    1-F = (n-1)(n-2) \frac{\pi^6}{16 t^4\Omega_c^4}\text{.}
\end{equation}

Note that the gate time to keep constant fidelity still scales as $t = O\left(\sqrt{n}\right)$. This fidelity scales worse with the number of qubits than the previously computed upper bound. We therefore expect this approximation to become invalid for high numbers of qubits, when this approximation becomes larger than the upper bound. This indicates that in the limit of a high number of qubits, the timing condition for the off-resonant excitations will become ineffective, and we are left with the fidelity given by Eq. \eqref{eq:4-level-phonon-fidelitybound}.

\section{Numerical Analysis}\label{sec:numericalanalysis}
We will now simulate the gate to demonstrate the working principles, show agreement with the theory, and assess performance. To this end, we used the Python package QuTiP \cite{johanssonQuTiP2Python2012}. 

\begin{figure}[h]
    \centering
    \includegraphics[width=\linewidth]{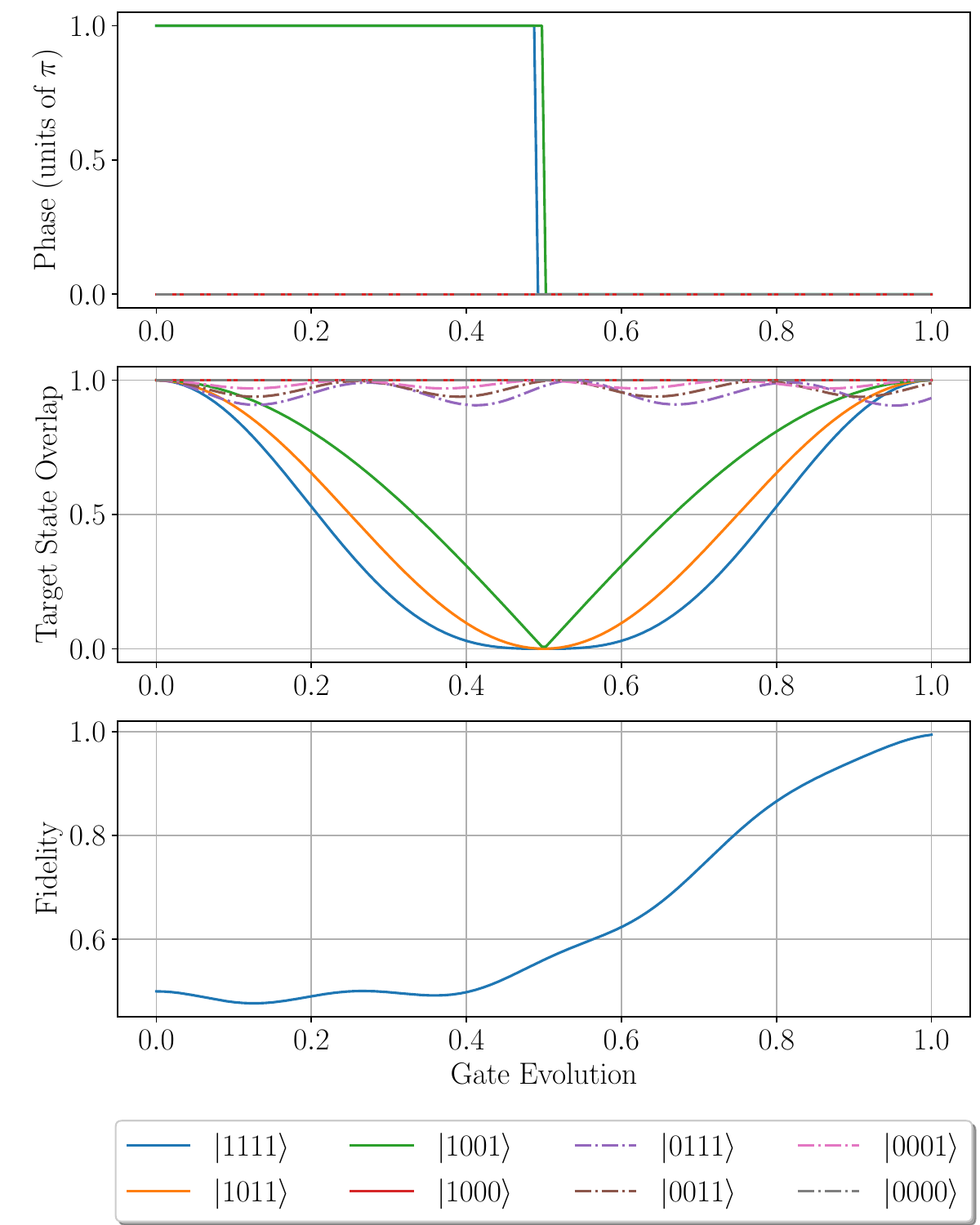}
    \caption{Numerical simulations of the scheme with three target qubits for different computational basis states. The drive ratio was chosen as $\Omega_t / \Omega_s = 1 / 8$. The first and second plot shows the amplitude and phase of the target state overlap, i.e. $\langle \psi_\text{target} | \psi(t)\rangle$ where $\ket{\psi_\text{target} }$ is the target state under the fanout gate, and $\ket{\psi(t)}$ its evolution under the Hamiltonian.The third plot shows the gate fidelity. The final fidelity is found to be $F=0.994$.}    \label{fig:4-level-phonon-simulation}
\end{figure}
\subsection{Demonstrations of the 4-Qubit Fanout Gate}
This scheme is performed in multiple stages. First, an excitation is produced in the harmonic oscillator using a Jaynes-Cummings $\pi$ pulse, then we apply resonance engineering to the 1-excitation manifold to apply $Z$ gates to the target qubits conditioned on the phonon occupation, and then the excitation is removed again. Without dissipation or thermal occupation in the harmonic oscillator, only the middle step (described by the Hamiltonian $H_2$ in Eq. \eqref{eq:H_2}) is subject to errors. We will therefore concentrate our analysis on this step.

The evolution of the basis states is shown in Fig. \ref{fig:4-level-phonon-simulation}. The first two plots show the phase $\phi$ and amplitude $A$ of the overlap 
\begin{equation}
    \langle \psi_\text{target} | \psi(t)\rangle \equiv A(t) e^{i\phi(t)}
\end{equation}
of the state with its target state under the fanout gate. Under the ideal gate operation, the amplitude goes to $1$ and the phase goes to $0$. The states in the transition subspace (control qubit $\ket 1$) -- shown with solid lines -- are taken through excited states and acquire a phase of $(-1)^m$, where $m$ is the number of target qubits in the $\ket 1$ state. During this transition, the amplitude in the second plot goes to zero, at which point a phase flip occurs. This is the same operating mechanism as the Cirac-Zoller gate \cite{ciracScalableQuantumComputer2000}: transferring the population through an excited state yields a phase factor of $-1$.

The states in the idle subspace (control qubit $\ket 0$) are shown using dashed lines. These states are in the $1$-excitation manifold and remain idle. They are subject to off-resonant excitations, which occur at approximately the same frequency for all states. This allows us to implement the timing condition mentioned in the previous chapter, such that the off-resonant excitation disappears. We reduce the strength of the strong anti-Jaynes-Cummings drive $\Omega_c$ until the timing condition \ref{eq:4lp_timing_condition} is met.

\begin{figure}
    \centering
    \includegraphics[width=\linewidth]{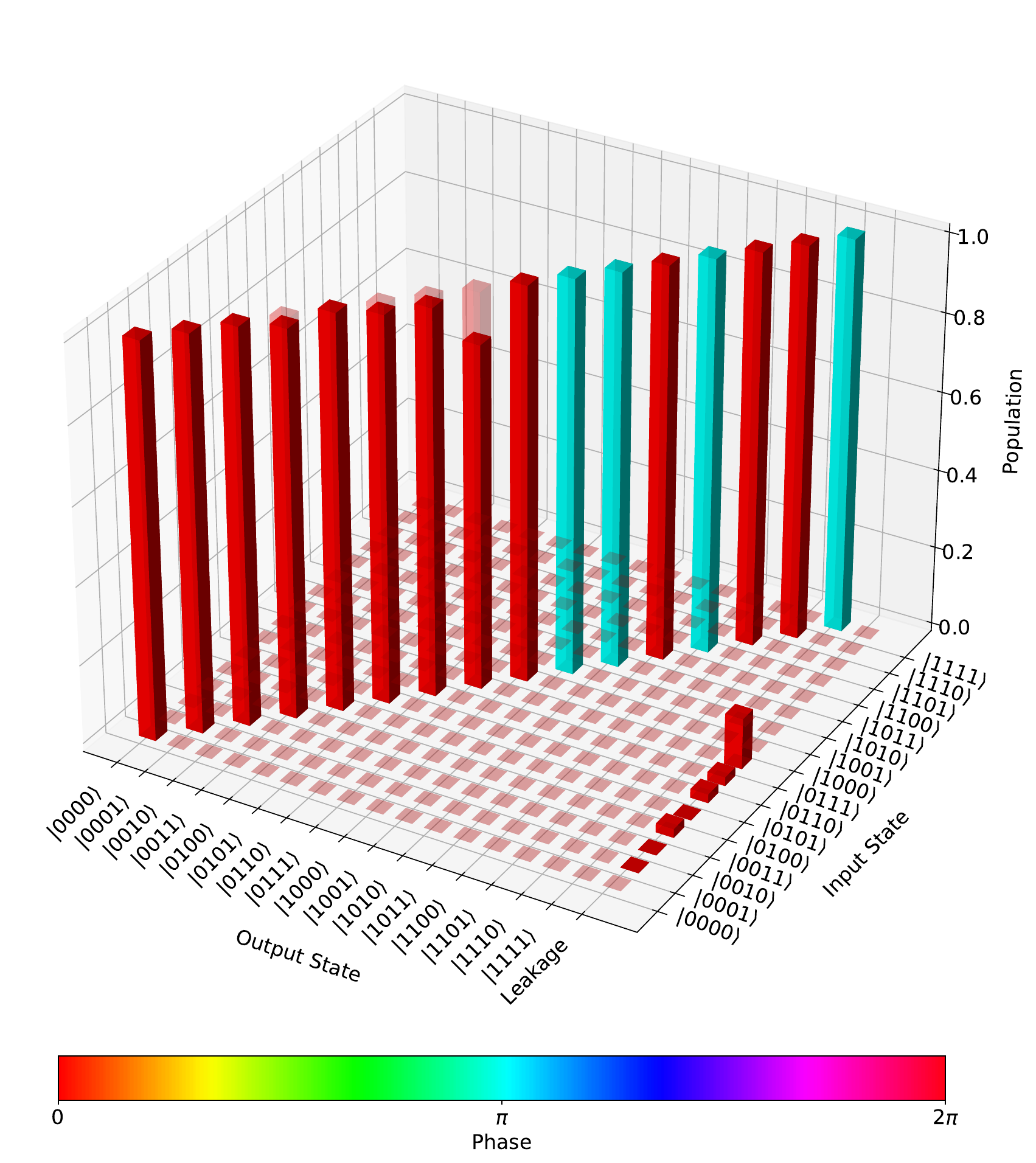}
    \caption{An input-output mapping of computational basis states. The height of the plot represents the populations in a specific target state, and the color shows the phase it acquires, as given by the color bar on the right. We observe that the evolution is indeed ideal on the transition subspace, and subject to off-resonant excitations into an excited state (labeled \textit{leakage}). The magnitude of the off-resonant excitations depends on the  number of target qubits in the $\ket 1$ state.}
    \label{fig:4-level-phonon-truthtable}
\end{figure}

The mapping from input states to output states under the gate evolution is shown in Fig. \ref{fig:4-level-phonon-truthtable}, for the case of three target qubits. The height of the bars show the population, and the color of the bar corresponds to the phase, meaning this plot is essentially a graphical representation of the unitary gate operator 
\begin{equation}
    U = \prod_{i}\exp(-iH_i t_i) 
\end{equation}
on the computational subspace. Fig. \ref{fig:4-level-phonon-truthtable} demonstrates that the gate acts ideally on the transition subspace (control qubit $\ket{1}$), and is subject to small off-resonant excitations into an auxiliary state on the idle subspace  (control qubit $\ket{0}$). 

\begin{figure}
    \centering
    \includegraphics[width=\linewidth]{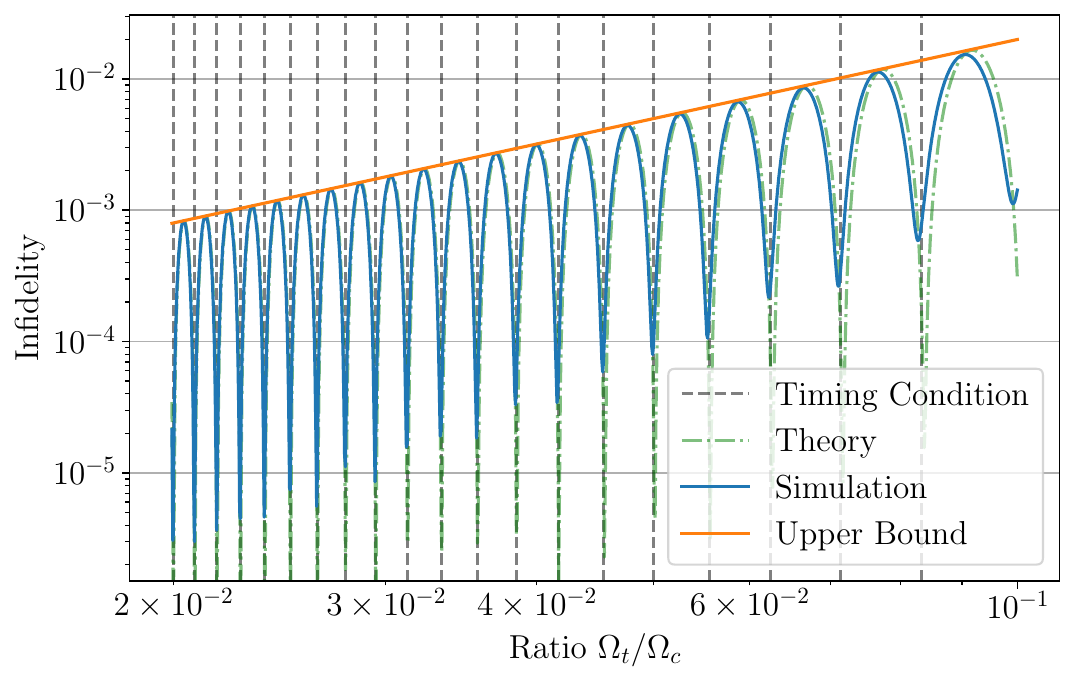}
    \caption{The gate infidelity as a function of the ratio between the weak $\Omega_t$ and strong $\Omega_c$ drives for a fanout gate with two target qubits. We observe good agreement between the simulated fidelity (blue), theoretical fidelity (green) the upper bound for the infidelity (orange) calculated in the previous section (Eq. \eqref{eq:4-level-phonon-fidelitybound}). The grey dashed lines are the ratios for which the timing condition is satisfied. When $\Omega_t$ becomes comparable to $\Omega_c$, the simulated fidelity deviates from the theoretical result. This is due to higher order excitations of the probe drive which were neglected in the theoretical analysis, and truncating terms of order $O\left(\frac{\Omega_t^4}{\Omega_c^4}\right)$.}
    \label{fig:4-level-phonon-fidelity}
\end{figure}

Now that we have illustrated the working mechanism of the gate, we would like to investigate the fidelity. We begin by simulating the ideal gate for different ratios $\Omega_t/\Omega_c$. The resulting plot is shown in Fig. \ref{fig:4-level-phonon-fidelity}. We observe that the simulated fidelity agrees with the upper bound calculated in Sec. \ref{sec:4-level-phonon-fidelity}. 

For lower gate times, the drive $\Omega_t$ becomes comparable to the the blocking drive $\Omega_c$. This results in a deviation of the theory from the simulation, as we assumed that $\Omega_t \ll \Omega_c$ in two steps of our analysis: We assumed that there are no second order excitations from the probe drive $\Omega_t$ (see Fig. \ref{fig:4-level-phonon_combined} b), and we truncated terms of order $O\left(\frac{\Omega_t^4}{\Omega_c^4}\right)$ in several steps of the analysis.

\subsection{Exact Simulation Up To 100 Qubits}
In our theoretical analysis, we simplified the system by truncated higher order drives. This simplification is justified by the fact that in the limit $\Omega_t \ll \Omega_c$, the higher states will not be populated. This allows us to derive an accurate upper bound for the gate error. As seen in Fig. \ref{fig:4-level-phonon-fidelity}, this simplification is valid for $\Omega_t/\Omega_c \rightarrow 0$, but results in a noticeable shift in the optimal timing condition for larger values of $\Omega_t/\Omega_c$. We therefore investigate the full system, including all the higher order drives in our numerical analysis. 

Simulating a full Hamiltonian to large numbers of qubits is typically infeasible due to the exponential dimensionality of the Hilbert space. However, by rewriting the Hamiltonian in terms of Dicke states \cite{dickeCoherenceSpontaneousRadiation1954}, we observe that the computational subspace only couples to a subset of states that grows quadratically with the number of qubits. This allows the exact simulation of our ideal gate operation to 100 qubits, as we will demonstrate below.

\begin{figure}[h]
    \centering
    \includegraphics[width=\linewidth]{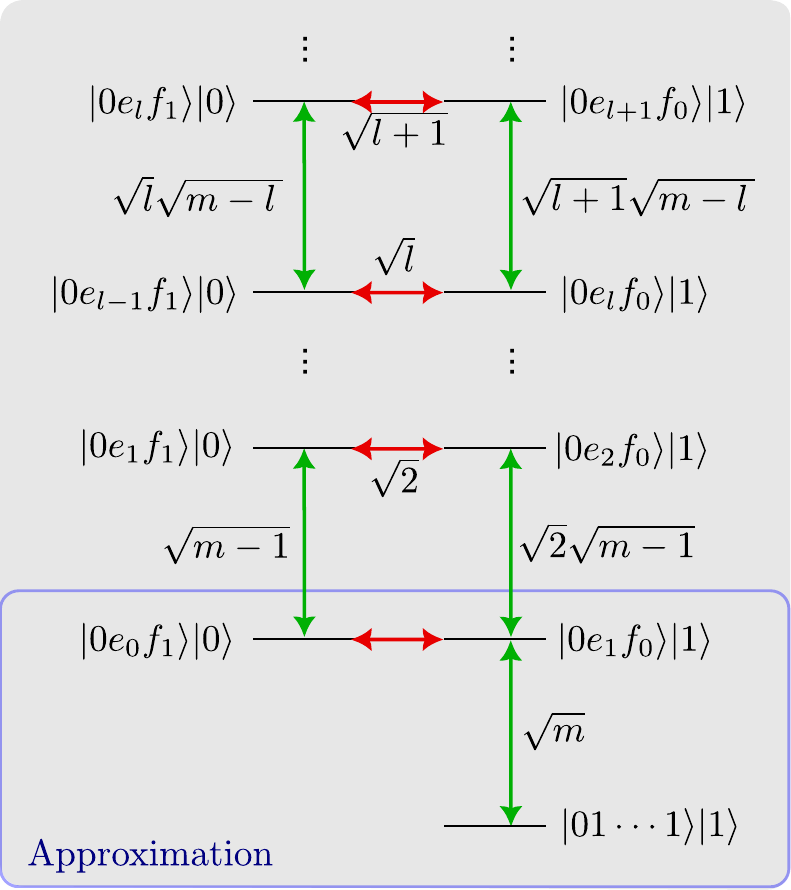}
    \caption{The couplings present on a state with $m$ target qubits in the state $\ket 1$, expressed in terms of Dicke states. The notation $\ket{0e_{n_e}f_{n_f}}\ket{k}$ represents an equal superposition of states with control qubit $\ket 0$, $n_e$ target qubits in the state $\ket e$ and $n_f$ target qubits in the state $\ket f$, with $k$ excitations in the harmonic oscillator. The simplification in the theoretical section only considered the lowest 3 states of this diagram, as indicated by the box.}
    \label{fig:couplings}
\end{figure}

In the Dicke basis (see App. \ref{app:dickebasis}), the generating Hamiltonian of our gate (Eq. \eqref{eq:H_2}) is block-diagonal: Each computational basis state lies in its own block, with couplings as shown in Fig. \ref{fig:couplings}.

Notice that the couplings in Fig. \ref{fig:couplings} only depend on the number of target qubits $m$ in the $\ket 1$ state. This allows us to recycle the computation of identical blocks, further reducing computational complexity. Instead of simulating a system of dimension $d=4^n$, we can simulate $n$ blocks of Hilbert space dimension $d_m=2m+1$, were $m$ is the number of target qubits in the state $\ket 1$. 

\begin{figure*}
    \centering
    \includegraphics[width=\linewidth]{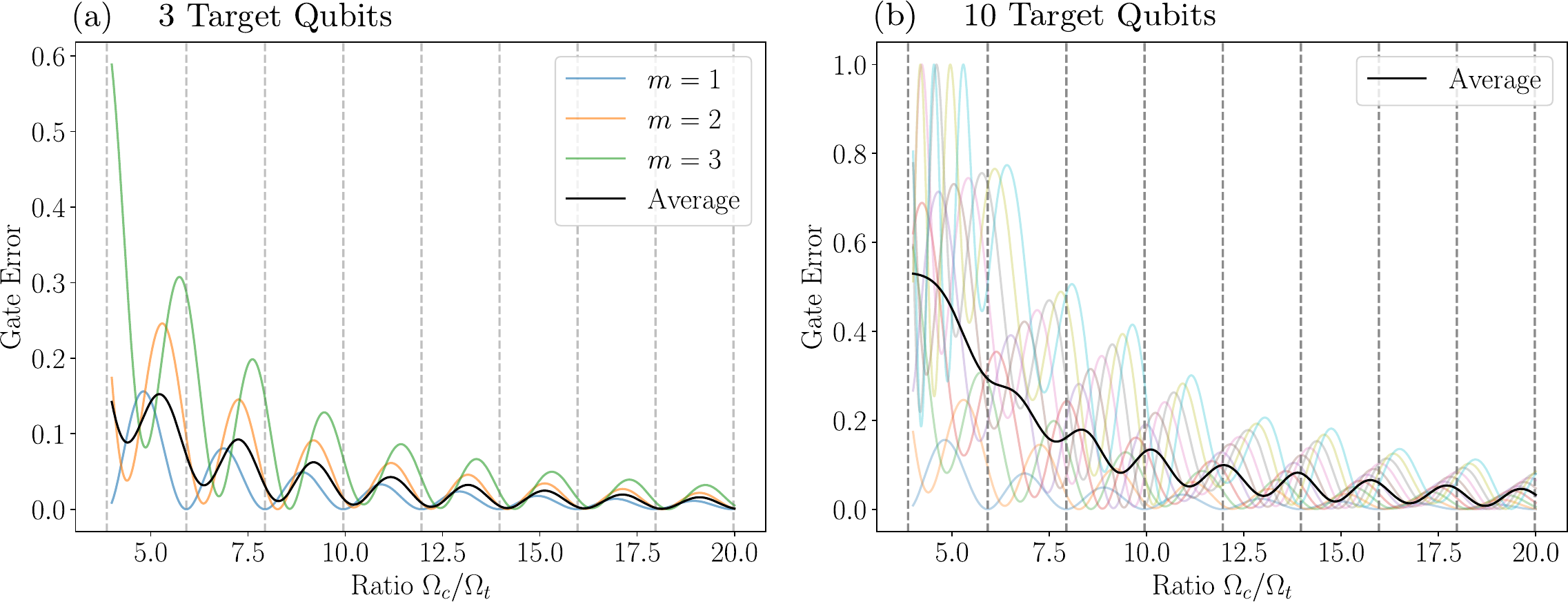}
    \caption{The error probability for different computational basis states in the idle subspace, in the case of low and higher numbers of qubits. The variable $m$ refers to the number of target qubits in the state $\ket 1$. 
    (a) The error probability for up to three target qubits. We observe periodic behavior as predicted in the theory section, with the theoretical optimum shown as the vertical dashed lines. Due to the inclusion of higher order excitations, the optimal timing condition has a shift which depends on $m$. The resulting average fidelity over all computational basis states has a shift in the timing condition derived in the theoretical section. We observe that for a high ratio $\Omega_c / \Omega_t$, this shift is lower and the simplification in the theory becomes justified. 
    (b) The error probability for 10 target qubits. We observe that for low $\Omega_c / \Omega_t$, the timing condition becomes ineffective as the minima disappear, and it  simply becomes optimal to choose the largest possible $\Omega_c$.}
    \label{fig:timing_condition_both_m}
\end{figure*}

This reduces the computational complexity from $O\left(\exp(n)\right)$ to $O(n^2)$ -- an exponential improvement which will allow us to simulate the system to high numbers of qubits. Note that the Dicke basis representation is exact -- we did not truncate or approximate the system in any way, but merely chose a different basis to represent the Hamiltonian.

Using this representation, we can simulate the off-resonant excitation for each computational basis state in the idle subspace. We consider the error probability
\begin{equation}
    1-|\bra{01\cdots1} e^{-iH_m t} \ket{01\cdots1}|^2\text{,}
\end{equation}
where $H_m$ is the Hamiltonian corresponding to Fig. \ref{fig:couplings}. 

We plot the resulting error probability as a function of the ratio $\Omega_c / \Omega_t$ in Fig. \ref{fig:timing_condition_both_m} (a). Comparing with the lowest order approximation, given by the vertical lines in the plot, we observe a shift in the timing condition due to the inclusion of higher order excitations. This shift depends on the variable $m$. This shift becomes smaller for larger $\Omega_c / \Omega_t$, which is expected from the theory: The larger the drive $\Omega_c$, the less populated the higher excited states will be. We observe that despite this shift, it is still possible to choose an optimal value for $\Omega_c / \Omega_t$ which minimizes the error due to off-resonant excitations.

Now we will consider the system in the limit of larger $m$. It is expected that the approximation made in the theory breaks down for large $m$, since the off-resonant excitations to higher order states are suppressed if $m\Omega_t \ll \Omega_c$. In Fig. \ref{fig:timing_condition_both_m} (b) we observe that for lower $\Omega_c / \Omega_t$, the shifts in the timing condition between different computational basis states become larger than the periodicity of the oscillations themselves. Averaging over all computational basis states, this renders the timing condition ineffective in the limit of large systems. Furthermore, the inclusion of higher order excitations causes a shift in the optimal timing condition from the theoretical prediction. One should therefore simulate the system numerically and minimize the error by varying $\Omega_c$ by the oscillation frequency of the off-resonant excitations. This frequency was found to be $2\Omega_t$ in the theoretical analysis. 

\begin{figure*}
    \centering
    
    \includegraphics[width=\linewidth]{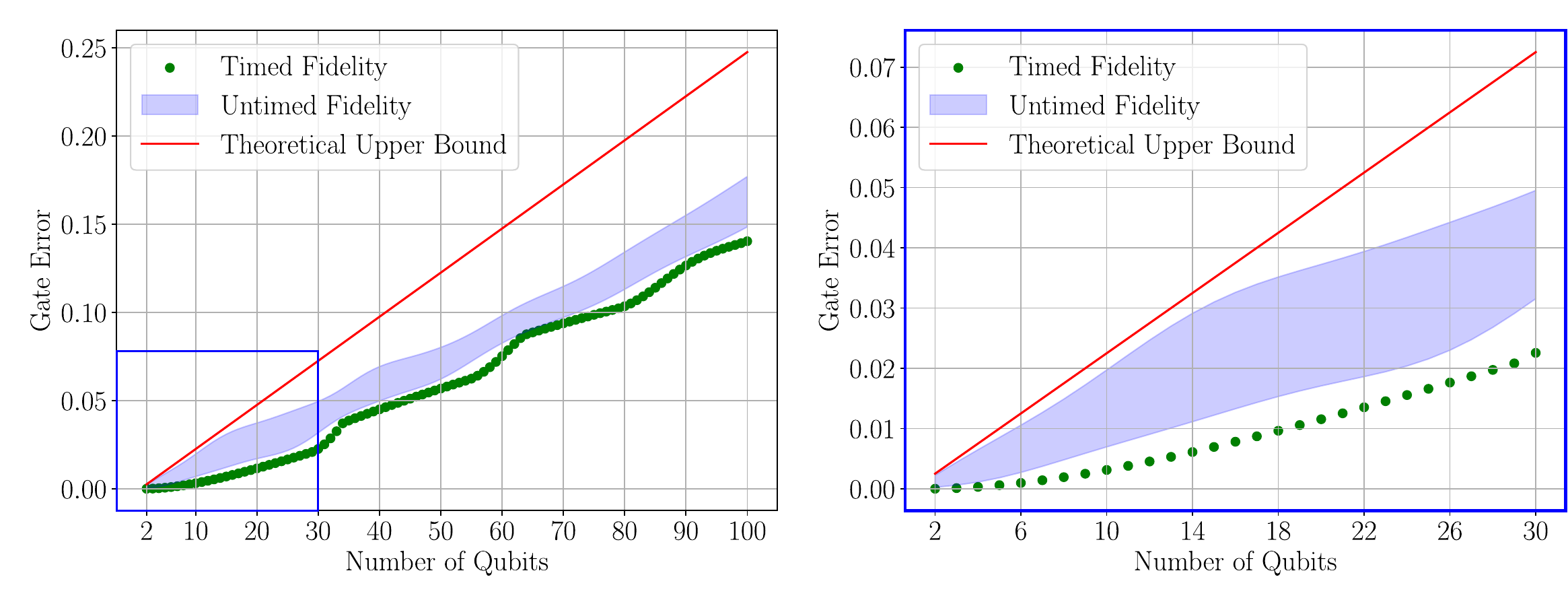}

    \caption{The simulated fidelity for up to 100 qubits (a) and a zoomed range up to 30 qubits (b), assuming a ratio $\Omega_s / \Omega_t \leq 20$. For the timed fidelity (indicated by the green dots), we choose the value within the interval $[18,20]$ which minimizes the gate error. The untimed fidelity shows the mean gate error for a sample of parameters within this interval. This simulates the expected performance if we ignored the timing condition. The red line is the theoretical upper bound given by Eq. \eqref{eq:4-level-phonon-fidelitybound} calculated in the theory section.}
    \label{fig:fidelity_scaling_R20_4lp}
\end{figure*}
We simulate the fidelity for different numbers of target qubits and analyze how the operation scales, with the resulting fidelities shown in Fig. \ref{fig:fidelity_scaling_R20_4lp}. To determine the fidelity for an optimally timed gate, we simulate each with slightly different strong sideband strengths $\Omega_c$ and choose the maximum. The simulations verify the validity of the upper bound while also demonstrating the effectiveness of the timing condition for up to 100 qubits, assuming we can achieve a ratio $\Omega_c/\Omega_t = 20$.

We observe that the fidelity scales linearly and respects the upper bound predicted in the theoretical section. For high qubit numbers, the timing condition becomes ineffective as predicted by the theory. The bumps in the optimally timed fidelity occur when the optimal $\Omega_c$ is on the boundary of the interval $\Omega_s/\Omega_t \in [18,20]$, requiring a discrete jump from one minimum to the next. The simulations demonstrate the validity of the analytical upper bound derived using the simplified system in the theoretical section.

\subsection{Performance under Heating}
We analyze the effect of heating of the harmonic oscillator on the gate fidelity. Such dissipative systems are more computationally expensive to simulate, as an exact simulation requires keeping track of density matrices rather than wavefunctions, which have quadratically more entries. As the number of states already grows exponentially with the number of qubits, this quickly becomes infeasible. In this section, we will use the permutation symmetry of the target qubits to reduce the exponential Hilbert space dimension to blocks of size $O(n^2)$, in a similar manner to the block-diagonalization of the pure state simulation. We then estimate the gate fidelity using quantum trajectory simulations \cite{dalibardWavefunctionApproachDissipative1992}.

We will use the heating operators for a harmonic oscillator in the limit of a hot bath, which are given by
\begin{equation}
    c_\text{heating} = \sqrt \kappa a^\dagger,\quad c_\text{cooling} = \sqrt \kappa a\text{,}
\end{equation}
where $\kappa$ is the heating rate. 

We use the fact that the Hamiltonian, as well as the dissipation operators respect a block-diagonal structure in the Dicke basis. In App. \ref{app:blockdiagonal-liouvillian}, we show that therefore the resulting quantum Master equation can also be described as an independent evolution for each of the blocks.

Each of the computational basis states $\ket {b_0 b_{1}\cdots b_n}\ket{p}$ lies in its own block. Here $b_i\in \{0,1\}$ are the qubit states and $p$ is the phonon occupation. In the Dicke basis defined in the earlier sections, we can express the action of the Hamiltonian \ref{eq:H_2} within each block as
\begin{align*}
    H &= \sum_l \sum_k \sum_p \Big(\Omega_t \sqrt{(l+1)(m-l)}\ketbra{e_{l+1}f_k,p}{e_l f_k, p} \\
    &+ \Omega_s \sqrt{(k+1)(l-k)}\sqrt p \cdot \ketbra{e_{l-1}f_{k+1},p-1}{e_l f_k, p} \Big)\\
    & + \text{H. c.}
\end{align*}

The numbers $m$, $l$ and $k$ denote the number of target qubits which are in the state $\ket 1$, $\ket e$ and $\ket f$ respectively. Note that for brevity, we are dropping additional indices to denote the computational basis state, as the couplings on each block only depend on the number of target qubits in the state $\ket 1$, denoted $m$.

By identifying $\ket {b_1\cdots b_n}\ket p = \ket{e_0 f_0}\ket{p}$, we can generate a similar energy level diagram as Fig. \ref{fig:couplings} for general phonon occupations.

We can now also express our collapse operators in this basis, by recognizing that the dissipation operators acting on the harmonic oscillator cause no cross-coupling between the different blocks. We can thus express the annihilation operator as
\begin{equation}
    a = \sum_l\sum_k\sum_p\sqrt{p}\ketbra{e_l f_k,p-1}{e_l f_k, p}\text{.}
\end{equation}

By using the fidelity approximation in Eq. \eqref{eq:approx_fidelity}, we can now simulate the fidelity of each of the computational basis states and compute the total fidelity. This simplification again ignores phase shifts between the blocks, and thus simulating dephasing operators in this manner is not possible. 

\begin{figure}
    \centering
    \includegraphics[width=\linewidth]{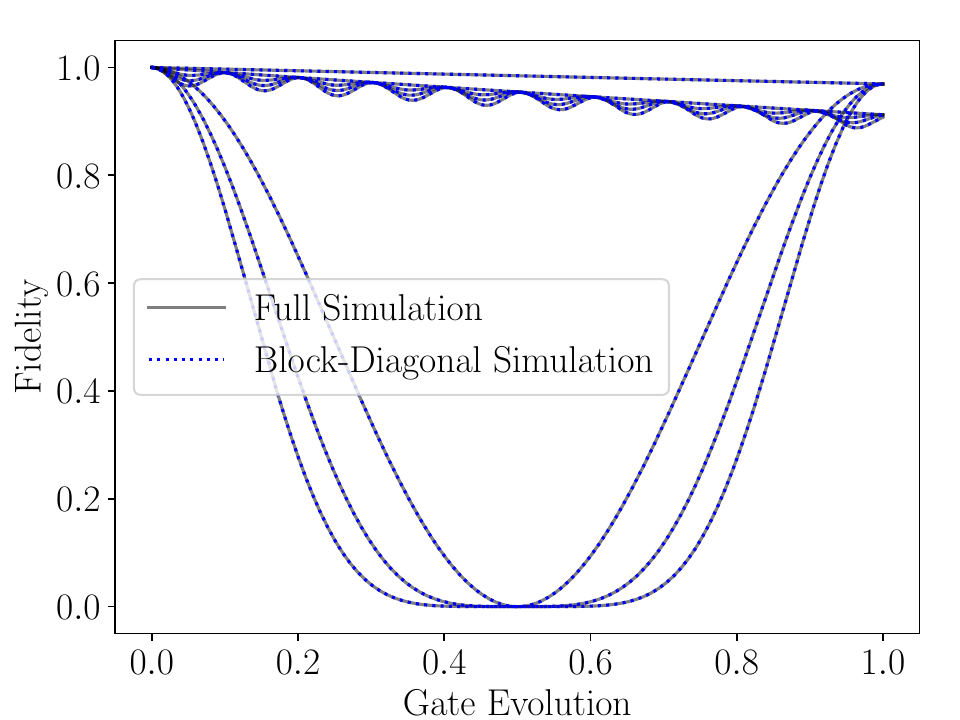}
    \caption{The fidelity of the 4-qubit fanout gate for different computational basis states with their target states, for the exact Hamiltonian and block-diagonal approximation. We assume $\Omega_s / \Omega_t \approx 20$ chosen optimally to match the timing condition of the gate, and a heating rate of $\kappa = 0.01\Omega_t$.}
    \label{fig:verify_basis_vectors_blockdiagonal}
\end{figure}

We would now like to verify the validity of our approximation. We first consider the simulation of computational basis states. This is to verify the correct implementation of the block-diagonalization. In Fig. \ref{fig:verify_basis_vectors_blockdiagonal}, we show the fidelity of each computational basis state with their target state, for the exact and block-diagonal evolutions. We observe perfect agreement between the two simulations, indicating a correct implementation of the block-diagonal Hamiltonian.

\begin{figure}[h]
    \centering
    \includegraphics[width=\linewidth]{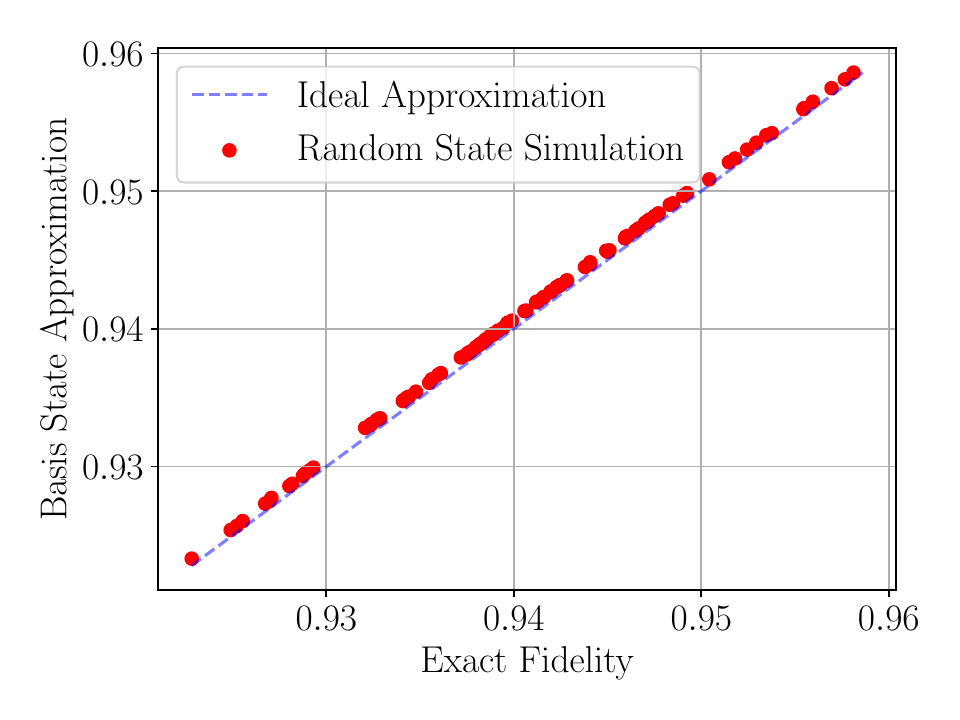}
    \caption{A comparison of the basis state approximation for the fidelity (Eq. \eqref{eq:approx_fidelity}) and the exact average gate fidelity, for 100 different random basis vectors for the $4$-qubit fanout gate. In this plot, we assumed $\Omega_s / \Omega_t \approx 20$ to be chosen to match the timing condition and a heating rate $\kappa = 0.01 \Omega_t$. The RMSE of the approximation was found to be 0.063\%, indicating excellent agreement.}
    \label{fig:test_heating_scatterplot}
\end{figure}
\label{app:blockdiagonal-liouvillian-verification}
We noted that the approximation of the fidelity in terms of basis vectors, as given by Eq. \eqref{eq:approx_fidelity}, is subject to errors if the different basis vectors acquire relative phases. This now also includes potential sources of dephasing. We therefore  compare the final fidelity of randomly chosen vectors using the exact and approximated fidelity. The comparison for the final fidelity of different random state vectors is shown in the scatter plot shown in Fig. \ref{fig:test_heating_scatterplot}.

The comparison demonstrates excellent agreement, with a Root Mean Squared Error (RMSE) well below 0.1\%. Notably, the basis state approximation always overestimates the fidelity. This is not surprising, as the approximation ignores errors due to dephasing between different blocks.

\begin{figure}
    \centering
    \includegraphics[width=\linewidth]{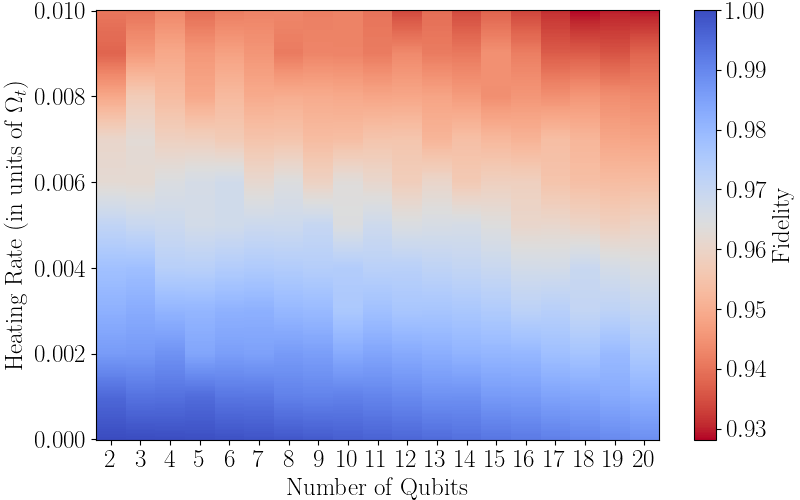}
    \caption{The fidelity for fanout gates with various qubit numbers for different heating rates of the harmonic oscillator, simulated using the diagonalization described in the text using the quantum trajectory method to integrate the dynamics. The ratio between the drives is chosen optimally in the interval $\Omega_c/\Omega_t \in [18,20]$.}
    \label{fig:heating_plot}
\end{figure}

The above block-diagonalization drastically reduces the number of states that need to be considered, and allows us to simulate the fidelity including the heating with up to $20$ qubits. This resulted in the plot shown in Fig. \ref{fig:heating_plot}. The simulated fidelities shown are the result of quantum trajectory simulations with 1000 trajectories, and thus results in some variance on the fidelity. 

We observe that the infidelity due to heating does not increase significantly with the number of qubits. The increase in infidelity for higher numbers of qubits in Fig. \ref{fig:heating_plot} is likely due to coherent errors and not dissipative errors.

\subsubsection{Experimental Comparison}
To estimate the feasibility of our scheme, we interpret our results here with realistic parameters from the Penning trap at the Trapped Ion Quantum Information group at ETH Zurich \cite{jainPenningMicrotrapQuantum2024}. Their system has a sideband strength of
\begin{align}
    \Omega_s = \eta \Omega \approx 20\text{ kHz} \cdot 2\pi\text{,}
\end{align}
where $\eta=0.4$ and $\Omega=50 \text{ kHz} \cdot 2\pi$ are the Lamb-Dicke parameter and Rabi rate respectively. Their heating rates are as low as $1 \text{ phonon}/\text{s}$, which results in a heating rate of
\begin{equation}
    \kappa = 1\text{ Hz} = 0.001 \Omega_t / 2\pi
\end{equation}
if we choose $\Omega_s / \Omega_t = 20$ as is done in Fig. \ref{fig:heating_plot}. Referring to plot the plot, we observe that due to the systems low heating rates, their system would incur no significant errors due to heating.

\section{Conclusion and Outlook}
In this article, we demonstrate a novel approach to implementing a many-qubit fanout gate based on resonance engineering in a collectively coupled qubit–oscillator system. By creating an excitation conditioned on the control qubit state, and exploiting excitation-dependent blocking arising from Jaynes–Cummings interactions between the target qubits and a common harmonic oscillator, we implement a fanout gate at the system level.

Through a theoretical analysis, we proved lower bounds on the gate fidelity, demonstrating improved scaling of our fanout gate.
Our scheme offers a favorable trade-off compared to two-qubit gate decompositions of
$n$-qubit fanout gates, delivering constant-depth operation at the cost of linear infidelity scaling. This represents an advantageous exchange: while fidelity scales similarly to conventional methods, the circuit depth is dramatically reduced compared to a decomposition into two-qubit CNOT gates.


We analyze the performance of our gate for large system sizes by exploiting permutation symmetry to reduce simulation complexity. By choosing an appropriate basis to represent the dynamics, we reduced the number of states considered in our simulation from exponential to quadratic in the number of qubits. This allows for the exact simulation of up to 100 qubits, as well as simulations with heating up to 20 qubits. This numerical analysis is consistent with our analytical findings for all system sizes considered.

From a methodological standpoint, our work provides a concrete example of how resonance engineering can be used to construct scalable quantum gates with well-characterized performance. While focused on a specific application, the approach developed here is broadly applicable and may inform the design of other multi-qubit operations, including dissipation engineering \cite{rojkovScalableDissipativeQuantum2025,reiterDissipativeQuantumError2017}, dissipative ground state preparation \cite{cubittDissipativeGroundState2023} and non-unitary gate implementations \cite{vanmourikExperimentalRealizationNonunitary2024}. 

In the context of discrete value error correction, stabilizer readouts can be expressed as a fanout gate. Here, the improved performance compared to a CNOT decomposition could lead to faster cycle times and higher fidelity operations, bringing us closer to fault-tolerant quantum computation \cite{googlequantumaiandcollaboratorsQuantumErrorCorrection2025}. 
The advantages extend to the logical level: because our fanout gate maintains transversality in any quantum error correcting code where the CNOT is transversal, it can be seamlessly integrated at the logical level, ensuring that the performance benefits will carry forward into the fault-tolerant era \cite{MultipleparticleInterferenceQuantum1996,gottesmanIntroductionQuantumError2010}.
\bigskip
\section*{Acknowledgements}
The authors would like to thank Jonathan Home and his research group for their support throughout the development of this work. We are grateful to Ivan Rojkov and Frederik van der Brugge for helpful discussions on theoretical aspects, and to Matteo Simoni and Pavel Hrmo for their insights on experimental aspects.
We acknowledge funding from the Swiss National Science Foundation (Ambizione grant no. PZ00P2\_186040) and the ETH Research Grant ETH-49 20-2.

\section*{Code Availability}
The code relating to the project is hosted on Github, available at \url{https://github.com/aljaeger/fanout}.


\bibliography{references-zotero.bib}

\onecolumngrid
\clearpage
\twocolumngrid

\appendix

\section{Canceling Off-Resonant Drives}
\label{sec:cancelling_raman}
Generally, an off-resonant drive causes an AC Stark shift on the states it couples. In the context of resonance engineering: The dressed states remain very close to the original states, but obtain an energy shift which depends on the coupling and detuning.

\begin{figure}[h]
    \centering
    \includegraphics[width=0.3\linewidth]{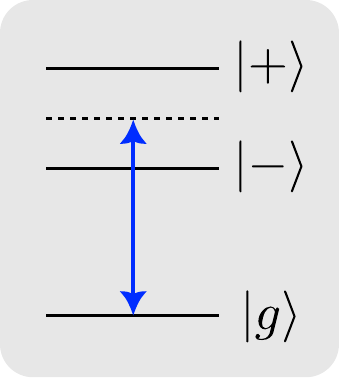}
    \caption{A double off-resonant drive to two levels with opposite detuning. Note that despite the fact that there is only one arrow in the diagram, these are in fact two separate drives.}
    \label{fig:doubleoffresonant}
\end{figure}

In this chapter, we will  consider the case of two off-resonant drives to two different auxiliary levels, with an opposite detuning, as shown in Fig. \ref{fig:doubleoffresonant}. As we will prove below, the two AC Stark shifts cancel each other in this case, and the only effect that remains is the (small) oscillation into the excited state. We will refer to this phenomenon as ``cancelling off-resonant drives''.

In the rotating frame, the system is described by the Hamiltonian 
\begin{align}
    H_{dr} = &+\Delta\ketbra{+}{+} - \Delta \ketbra{-}{-} \\
    &+\Omega(\ketbra{+}{0} + \text{H.c.}) \\
    &+\Omega(\ketbra{-}{0} + \text{H.c.})\text{.}
\end{align}

this can be expressed in matrix form as
\begin{equation}
    H_{dr} = \begin{pmatrix}
     0 & \Omega & \Omega\\
     \Omega & \Delta & 0\\
     \Omega & 0 & -\Delta
\end{pmatrix}\text{.}
\end{equation}

The time evolution of this Hamiltonian
\begin{equation}
    U(t) \equiv \exp(-i H t)
\end{equation}
has an analytic solution. We can compute that
\begin{align}
    \bra{g}U(t)\ket{g} &= \frac{\Delta^2 + 2\Omega^2\cos\left(t\sqrt{\Delta^2 + 2\Omega^2}\right)}{\Delta^2 + 2\Omega^2} \nonumber\\
    &= 1 - 4\frac{\Omega^2}{\Delta^2 + 2\Omega^2}\sin^2\left(\frac{t}{2}\sqrt{\Delta^2 + 2\Omega^2}\right)\text{.}
\end{align}

We notice that the state $\ket g$ does not obtain an AC stark shift, which would appear as a complex phase accumulation in the matrix element. It is only subject to off-resonant excitations of the order $O(\Omega^2/\Delta^2)$, which can be suppressed in the limit $\Omega^2 \ll \Delta^2$.

\section{Evaluation of Sums Using the Binomial Theorem}\label{sec:sum_calculation}
In this section, we show that
\begin{equation}
     \frac{1}{2^n}\sum_{m=0}^n \begin{pmatrix}
        n\\m
    \end{pmatrix}m = \frac{n}{2} \text{,}
\end{equation}
using the Binomial theorem. 

It follows from the binomial theorem that
\begin{equation}
    (x+1)^n = \sum_{m=0}^n \begin{pmatrix}
        n\\m
    \end{pmatrix}x^{m}\text{.}
\end{equation}
 We take the derivative with respect to $x$ and find
 \begin{equation}
    n(x+1)^{n-1} = \sum_{m=0}^n \begin{pmatrix}
        n\\m
    \end{pmatrix}m x^{m-1}\text{.}
\end{equation}
Substituting $x=1$, we find
\begin{equation}
    n 2^{n-1} = \sum_{m=0}^n \begin{pmatrix}
        n\\m
    \end{pmatrix}m \text{,}
\end{equation}
and therefore
\begin{equation}
     \frac{1}{2^n}\sum_{m=0}^n \begin{pmatrix}
        n\\m
    \end{pmatrix}m = \frac{n}{2} \text{.}
\end{equation}

Higher order sums, such as
\begin{equation}
    \frac{1}{2^n}\sum_{m=0}^n \begin{pmatrix}
        n\\m
    \end{pmatrix}m^2\text{,}
\end{equation}
can be evaluated similarly by taking more derivatives and using the previous results.

\bigskip
Interestingly enough, the same result can be proven using the symmetry of the binomial coefficient:
\begin{equation}
    \begin{pmatrix}
        n \\m
    \end{pmatrix} =\begin{pmatrix}
        n\\n-m
    \end{pmatrix}.
\end{equation}
By reversing the order of the sum, we find
\begin{align}
    \frac{1}{2^n}\sum_{m=0}^n \begin{pmatrix}
        n \\m
    \end{pmatrix}m &= \sum_{m=0}^n\begin{pmatrix}
        n\\n-m
    \end{pmatrix}(n-m)\\
    &=\sum_{m=0}^n\begin{pmatrix}
        n\\m 
    \end{pmatrix}(n -m).
\end{align}
We can rearrange this to 
\begin{equation}
     \frac{1}{2^n}\sum_{m=0}^n \begin{pmatrix}
         n\\m
     \end{pmatrix} m =  \frac{n}{2^n}\sum_{m=0}^n \begin{pmatrix}
         n\\m
     \end{pmatrix}  - \frac{1}{2^n}\sum_{m=0}^n \begin{pmatrix}
         n\\m
     \end{pmatrix} m \text{.}
\end{equation}
Using $\sum_{m=0}^n \begin{pmatrix}
         n\\m
     \end{pmatrix} =2^n$, dividing the equation by 2 and rearranging, we find the result
     \begin{equation}
         \frac{1}{2^n}\sum_{m=0}^n \begin{pmatrix}
         n\\m
     \end{pmatrix} m = \frac{n}{2}
     \end{equation}
\section{Representation of the Hamiltonian in Terms of Dicke States}
\label{app:dickebasis}
\begin{figure*}
    \centering
    \includegraphics[width=0.7\linewidth]{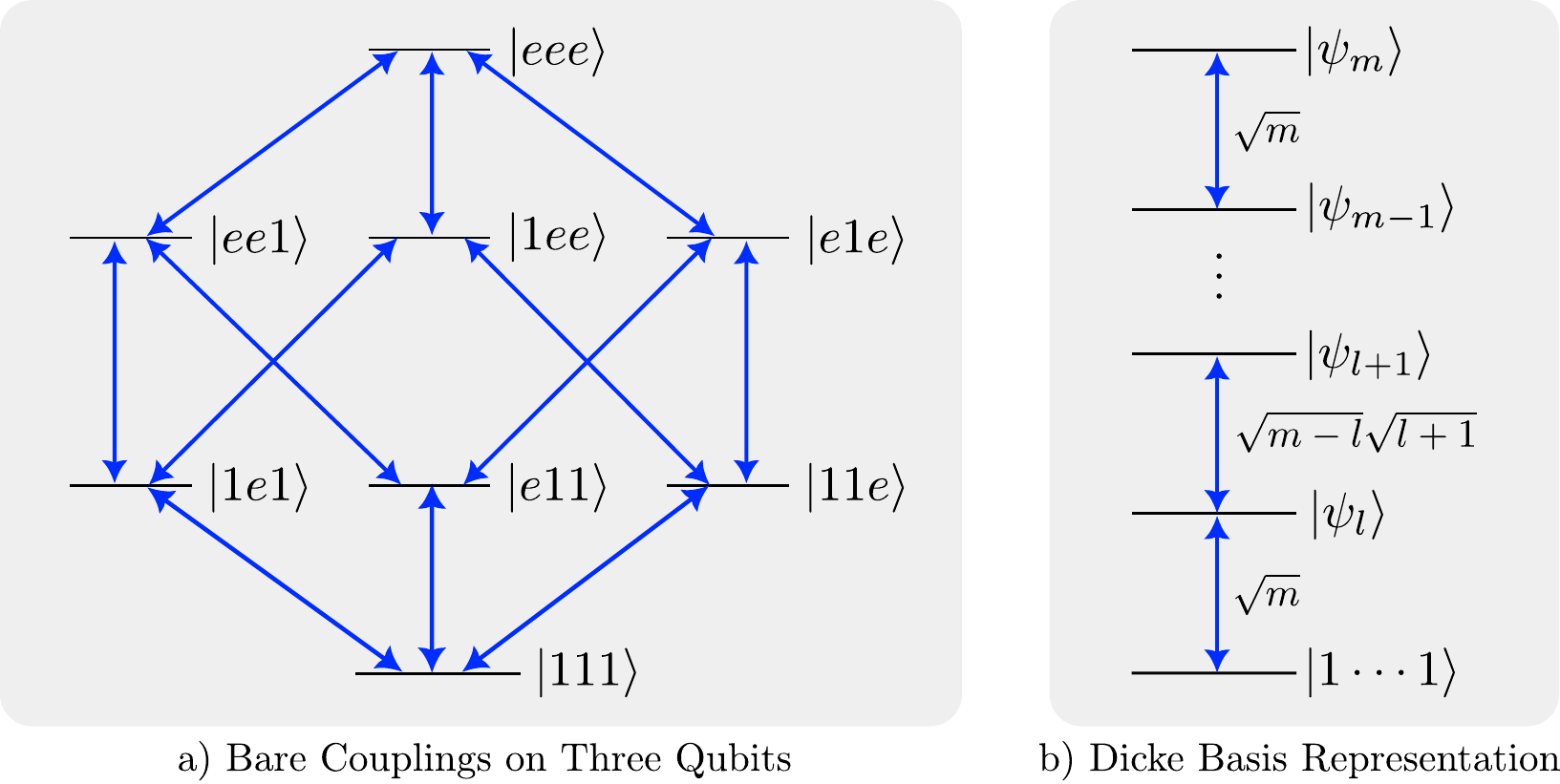}
    \caption{(a) The couplings present on the $\ket{111}$ due to local $\sigma_{1e} + \text{H.c.}$ interactions. These interactions can be expressed as collective couplings between Dicke states. The corresponding energy level diagram is shown in panel (b), generalized for a state with $m$ qubits in the $\ket 1$ state. The corresponding Hamiltonian is given by Eq. \eqref{eq:DickeLadder}. The Dicke states $\ket{\psi_l}$ in panel (b) are given by equal superpositions of states with $l$ qubits in the $\ket e$ state, i.e. all states in each level in panel (a). Expressing the Hamiltonian in this basis results in a collective enhancement of $\sqrt{m-l}\sqrt{l+1}$.}
    \label{fig:dickeladder}
\end{figure*}
The coupling on $n$ qubits can generally be expressed in terms of the Dicke basis, a number of collective excitations of the form
\begin{equation}
    \ket{k, N-k} = \sum_{\sigma \in S_n} \ket{b_{\sigma(1)},\cdots,b_{\sigma(n)}}
\end{equation}
where $S_n$ is the permutation group of $n$ elements and $b$ is a bit string with $k$ ones.

In this basis, we can express the operator 
\begin{equation}\label{eq:DickeLadder}
    \sum_i\sigma_x^i = \sum_k \sqrt{(N-k)(k+1)}\cdot \ketbra{k+1, N-k-1}{k, N-k}
\end{equation}

An energy level diagram for the above Hamiltonian is shown in Fig. \ref{fig:dickeladder}, showing the couplings in the computational basis for the case of 3 qubits in panel (a), and a general energy level diagram in the Dicke basis in panel (b).

Throughout the article, we use this representation but generalize it to more qubits by introducing a notation that tracks the number of excitations on each state $\ket{1_me_lf_k}$, where often we leave out the $\ket 1$ state for brevity.

\section{Block-Diagonality of the Quantum Master Equation}
\label{app:blockdiagonal-liouvillian}
Suppose both the Hamiltonian $H$ as well as the collapse operators $c_k$ are block-diagonal, with the same blocks. 
This means that there exists a set of projectors $P_i$ such that
\begin{align}
    H &= \sum_i P_i H P_i, \quad \text{and}\\
    c_k &= \sum_i P_i c_k P_i\text{.}
\end{align}
The quantum Master equation is given by
\begin{equation}
    \dot\rho = -i[H, \rho] + \sum_k c_k \rho c_k^\dagger + \frac{1}{2}\{c_k^\dagger c_k, \rho\}\text{.}
\end{equation}
We will assume a state starts in the block $P_i$, meaning that $\rho = P_i \rho P_i$. Substituting this, along with the block-diagonality assumptions above into the above quantum master equation we find
\begin{align}
    \dot  \rho  &= \sum_j\Big( -i[P_j HP_j, P_i \rho P_i] + \sum_k P_jc_kP_j P_i\rho P_i P_jc_k^\dagger P_j\nonumber \\
    &+ \frac{1}{2}\{P_j c_k^\dagger c_k P_j, P_i\rho P_i\}\Big)
\end{align}

We note that for projectors, $P_i P_j = \delta_{ij} P_i$. Using this identity, we can simplify the above equation to
\begin{align}
    \dot \rho &= \Big( -i[P_i H P_i, P_i \rho P_i] + \sum_k P_ic_kP_i P_i\rho P_i c_k^\dagger P_i \nonumber \\
    &+ \frac{1}{2}\{P_i c_k^\dagger P_i P_i c_k P_i, P_i\rho P_i\}\Big)
\end{align}
indicating the restricted dynamics on this block are governed by the Hamiltonian and collapse operators
\begin{align}
    H_i &= P_i H P_i\\
    c_k^i &= P_i c_k P_i\text{.}
\end{align}

This is exactly what we do in the article. For each computational basis state, we only consider the Hamiltonian and collapse operators restricted to its block.

\end{document}